\documentclass[showpacs,aps,nofootinbib]{revtex4-2}
\emergencystretch=\maxdimen
\usepackage{amsfonts,amsmath,amssymb}
\usepackage{latexsym,color}

\usepackage[utf8]{inputenc}
\usepackage{graphicx, psfrag}

\def\qq{\qquad}
\def\cm{\hspace*{1cm}}
\def\inch{\hspace*{1in}}

\def\lal{&&{}}
\def\eq{Eq.\,}
\def\eqs{Eqs.\,}
\def\beq{\begin{equation}}
\def\eeq{\end{equation}}
\def\bear{\begin{eqnarray}}
\def\bearr{\begin{eqnarray} \lal}
\def\ear{\end{eqnarray}}
\def\earn{\nonumber \end{eqnarray}}

\def\nnn{\nonumber\\ \lal }
\def\nnnv{\nonumber\\[5pt] \lal }

\def\yyy{\\[5pt] \lal }

\def\e{{\,\rm e}}
\def\d{\partial}

\def\diag{\mathop{\rm diag}\nolimits}

\def\const{{\rm const}}
\def\Half{\dfrac 12}

\def\then{\ \Rightarrow\ }

\def\eqn#1{\eq\eqref{#1}}
\def\rf{\eqref}
\def\mn{_{\mu\nu}}
\def\MN{^{\mu\nu}}
\def\mN{_\mu^\nu}

\def\R{{\mathbb R}}
\def\Z{{\mathbb Z}}
\def\cF{{\mathcal F}}
\def\cL{{\mathcal L}}
\def\GR{general relativity}
\def\sph{spherically symmetric}
\def\ssph{static, spherically symmetric}
\def\bh{black hole}
\def\bhs{black holes}
\def\wh{wormhole}
\def\whs{wormholes}
\def\asflat{asymptotically flat}
\def\emag{electromagnetic}
\def\Scw{Schwarz\-schild}
\def\RN{Reiss\-ner-Nord\-str\"om}

\begin{document}

\title{Black bounces, wormholes, and partly phantom scalar fields}

\author{K.A. Bronnikov}
\affiliation{Center of Gravitation and Fundamental Metrology, VNIIMS, Ozyornaya ulitsa 46, Moscow 119361, Russia}
\affiliation{Peoples' Friendship University of Russia, ulitsa Miklukho-Maklaya 6, Moscow, 117198, Russia}
\affiliation{National Research Nuclear University "MEPhI", Kashirskoe shosse 31, Moscow 115409, Russia}

\date{\today}

\begin{abstract}
   {Simpson and Visser recently proposed a phenomenological way to avoid some kinds of space-time
   singularities by replacing a parameter whose zero value corresponds to a singularity (say, $r$) with the
   manifestly nonzero expression $r(u) = \sqrt{u^2 + b^2}$, where $u$ is a new coordinate, and 
   $b = \const >0$. This trick, generically leading to a regular minimum of $r$ beyond a \bh\ horizon 
   (called a ``black bounce''), may hopefully mimic some expected results of quantum gravity, and 
   was previously applied to regularize the \Scw, \RN, Kerr and some other metrics. In this paper it is 
   applied to regularize the Fisher solution with a massless canonical scalar field in \GR\ (resulting in 
   a traversable \wh) and a family of \ssph\ dilatonic \bhs\ (resulting in regular \bhs\ and \whs). These 
   new regular metrics represent exact solutions of \GR\ with a sum of stress-energy tensors of a scalar 
   field with nonzero self-interaction potential and a magnetic field in the framework of nonlinear 
   electrodynamics with a Lagrangian function $\cL(\cF)$, $\cF = F\mn F\MN$. A novel feature in the 
   present study is that the scalar fields involved have ``trapped ghost'' properties, that is, are phantom 
   in a strong-field region and canonical outside it, with a smooth transition between the regions. 
   It is also shown that any \ssph\ metric can be obtained as an exact solution to the Einstein equations 
   with the stress-energy tensor of the above field combination.   }
\end{abstract}	
\maketitle
  
\section{Introduction}

  There seems to be a common belief that quantum gravity effects inevitably suppress space-time singularities 
  existing in the solutions of classical gravity, replacing them with some regular phenomena. Though, different 
  approaches and models of quantum gravity, when translated to the classical language, lead to radically 
  different results, see, e.g., \cite{QG1, QG2, QG3, kelly20, achour20, kunst20, ash20a, bambi13}, and a 
  discussion in \cite{we20}. Among them are \bh-white hole transitions  \cite{QG2, QG3, kelly20, achour20},
  scenarios with the spherical radius tending to a nonzero constant at late times beyond an event horizon 
  \cite{kunst20}, quantum-corrected configurations without horizons \cite{bambi13}, etc.   
  This diversity may be treated as a consequence of a still uncertain nature of quantum gravity by now. 
  
  Therefore, of evident interest is the recent proposal made by Simpson and Visser (SV) \cite{simp18} to
  regularize a singular space-time metric by simply replacing its certain parameter (specifically, the spherical 
  radius $r$ in a \ssph\ metric), whose zero value corresponds to a singularity, with the simple expression 
  $r(u) = \sqrt{u^2 + b^2}$, where $u \in \R$ is a new coordinate introduced instead of $r$,\footnote
  		{The notation $r$ is kept in this paper for the spherical radius, $r = \sqrt{-g_{\theta\theta}}$, while 
  		radial coordinates different from $r$ are denoted by other letters to avoid confusion.}  
  and $b > 0$ is a regularization parameter. It looks as an easy way to imagine the possible effects of quantum 
  gravity within a classical space-time framework, leaving aside any quantization details. Besides, new geometries
  emerging in this way may be of interest by themselves.

  In \cite{simp18} this proposal was applied to the \Scw\ \bh\ solution, resulting in the globally regular metric
\beq                    \label{S-SV}
		ds^2 = \bigg(1 - \frac{2m}{\sqrt{u^2 + b^2}}\bigg)dt^2 
			- \bigg(1 - \frac{2m}{\sqrt{u^2 + b^2}}\bigg)^{-1} du^2 
			- (u^2 + b^2) (d\theta^2 + \sin^2\theta d\varphi^2),
\eeq  
  where the singularity at $r=0$ is replaced by a regular minimum of $r(u)$ at $u=0$, a sphere of radius $b$. 
  Depending on the relation between $b$ and the \Scw\ mass $m$, the geometry \rf{S-SV} can represent 
  a \wh\  with a throat at $u=0$ (if $b > 2m$, hence $g_{tt}(0) > 0$), a \bh\ with two horizons at 
  $u = \pm \sqrt{4m^2 -b^2}$ if $b < 2m$, and an extremal \bh\ with a single horizon at $u=0$ if $b =2m$.
  
  In the \bh\ case $b < 2m$, the hypersurface $u=0$, being a minimum of $r(u)$, is not a throat
  since $u$ is a temporal coordinate there: instead, $u=0$ corresponds to a bounce in one of the two scale 
  factors, $r(u)$, of a Kantowski-Sachs cosmology in the inner region of the \bh, and by SV suggestion such
  a minimum is called a {\it black bounce} \cite{simp18}. (The other scale factor in this Kantowski-Sachs 
  cosmology is $2m/\sqrt{u^2 + b^2}  - 1$, and it has a maximum at the same time instant $u=0$.)
  It can be noted that black bounces are a common feature of another class of space-times, black universes, 
  that is, \bhs\ that contain an expanding asymptotically isotropic cosmology beyond the horizon 
  \cite{bk-uni05, bk-uni06, bk-uni12}. Such solutions are naturally obtained with a phantom scalar field as a 
  source both in \GR\ (GR) and its extensions \cite{clem09, azreg11, trap18}.
  
  In the intermediate case of \rf{S-SV}, $b=2m$, the hypersurface $u=0$ is null, it is simultaneously a throat 
  and a \bh\ horizon, and it was suggested \cite{rahul22} to call it a {\it black throat}. 
  
  A similar SV regularization of the Reissner-Nordstr\"{o}m space-time was considered in \cite{franzin21}. 
  A large class of regular \bh\ and \wh\ space-times was constructed using the SV approach by Lobo et al. 
  \cite{lobo20}. The rich and diverse geometries obtained in this way have attracted much interest, and their 
  extensions with rotation were also studied \cite{mazza21, xu21, shaikh21}. Further relevant studies
  concerned gravitational wave echoes at possible black hole/\wh\ transitions, quasinormal modes of such
  space-times and gravitational lensing phenomena
  \cite{churilova19, yang21, guerrero21, tsukamoto21, islam21, cheng21, kb20, tsukamoto20, lima21,
  nascimento20, franzin22}.
  
  The possible presentation of regular \ssph\ SV-like space-times as solutions to the equations of GR with field
  sources was considered in \cite{rahul22} and \cite{canate22}. As shown in \cite{rahul22}, a large class of
  SV space-times can be obtained as exact solutions to the Einstein equations with a source in the form of a 
  self-interacting minimally coupled phantom scalar field combined with an \emag\ field in the framework of 
  nonlinear electrodynamics (NED), whereas a scalar field alone or NED alone cannot provide a necessary
  material source for an SV space-time. Another method of obtaining such sources was formulated in
  \cite{canate22}, along with some new examples of interest. The necessity of a phantom field in our 
  construction is evident due to a minimum of the spherical radius $r$ in SV metrics, while the presence 
  of a NED source is needed for obtaining the relevant form of the stress-energy tensor (SET). For SV
  regularizations of the \Scw\ and \RN\ solutions of GR, the explicit forms of scalar and NED sources were
  obtained in \cite{rahul22}, and the global structure diagrams for the regularized \RN\ metric with three 
  and four horizons were constructed. Some analogues of SV regularization applied to cosmology have 
  been recently considered in \cite{kam22}.
  
  The present study further extends that of \cite{rahul22}. First, it is shown that a combination of NED and a 
  minimally coupled scalar field is able to provide a source for {\it any\/} \ssph\ metric in the framework of GR.
  It turns out, however, that the corresponding scalar field, in general, cannot be only canonical (that is, with 
  positive kinetic energy) or only phantom (with negative kinetic energy), but has to change its nature from 
  one region to another. Such a situation, where a scalar is phantom only in a strong-field region and is 
  canonical elsewhere, acquired the name of a ``trapped ghost'' and was shown to lead to globally regular 
  space-times including \whs\ and \bhs\ \cite{kroger04, trap10, don11}. Even more than that, it turned 
  out that the transition surfaces between the canonical and phantom scalar field regions can play 
  a stabilizing role in \bh\ and \wh\ space-times \cite{trap17, trap18}.
  
  Second, we here consider SV-like regularizations for two families of singular solutions of GR: Fisher's  
  solution with a massless canonical scalar field \cite{fisher48} and a special subset of dilatonic \bh\ solutions
  with interacting massless scalar and \emag\ fields \cite{dil1, dil2, dil3, dil4}. 
  In both cases, the SV substitution is applied in the simplest possible way 
  ($x \mapsto \sqrt{u^2 + b^2}$) to the factor $x$ that produces a singularity at its zero value. 
  The scalar-NED sources for regularized versions of these space-times are found, and it turns out that 
  trapped ghost scalars are necessary for their GR description.
  
  As always when dealing with phantom fields (or partly phantom, as trapped ghosts) one can recall 
  the well-known problem of their potential instability at the quantum level. This problem is widely
  discussed, and it is often said that a phantom field can be an effective entity, originating from some extended
  theory of gravity, for example, from extra-dimensional degrees of freedom (see, e.g., \cite{nilles84,khve98}
  and a discussion in \cite{don11}) and should not be quantized. Anyway, as in many other studies, the 
  existence of phantom fields is here admitted as a working hypothesis, which is quite necessary as long as 
  there are throats or black bounces.
  
  The paper is organized as follows. In Section II it is shown how to obtain a scalar-NED source for an arbitrary 
  \ssph\ metric using a magnetic field solution of NED. Sections III and IV are devoted to finding and describing 
  scalar-NED sources for regularized Fisher and dilatonic space-times. Section V contains some 
  concluding remarks. The metric signature $(+\,-\,-\,-)$ is adopted, along with geometrized units
  such that $8\pi G = c = 1$.

\section{Scalar--NED sources for spherical space-times}

  Consider an arbitrary \ssph\ metric in the form
\beq 		\label{ds}
		ds^2 = A(x) dt^2 - \frac{dx^2}{A(x)} - r^2(x) d\Omega^2,\qq 
		d\Omega^2 = d\theta^2 + \sin^2\theta d\varphi^2,
\eeq  
  written in terms of the so-called quasiglobal radial coordinate $x$ \cite{BR-book}. This choice of the radial 
  coordinate is well suited for the description of any \ssph\ space-times including \bhs\ (where horizons appear 
  as regular zeros of $A(x)$ provided $r(x)$ is finite) and \whs\ (where throats appear as regular minima 
  of $r(x)$ provided $A(x) > 0$). Recall once more that if a minimum of $r(x)$ occurs in a region where 
  $A(x) > 0$, it is called a \wh\ throat, such a minimum in a region where $A(x) < 0$ is named a black bounce 
  by suggestion of \cite{simp18}, and if a minimum of $r(x)$ coincides with $A(x) =0$, we call it a black 
  throat \cite{rahul22}.
   
  Let us show that {\it any\/} (sufficiently smooth) metric \rf{ds} is a solution to the Einstein equations with a 
  SET $T\mN$ of a non-interacting combination of a minimally coupled scalar field $\phi$ 
  with a certain potential $V(\phi)$ and a nonlinear \emag\ field with the Lagrangian density $\cL(\cF)$,
  where $\cF = F\mn F\MN$, and $F\mn = \d_\mu A_\nu - \d_\nu A_\mu$ is the \emag\ field tensor.  
  
  To begin with, the most general form of the SET compatible with the metric \rf{ds} according to the 
  Einstein equations 
\beq 			\label{EE}
		G\mN \equiv R\mN - \Half \delta\mN R = - T\mN, 
\eeq  
  has a form that formally coincides with the SET of an anisotropic fluid,
\beq                         \label{T-flu}
		T\mN [{\rm fluid}] = \diag(\rho, - p_r, -p_\bot, -p_\bot), 
\eeq  
  where $\rho, p_r, p_\bot$ are the density, radial pressure and tangential pressure, respectively.  
  
  On the other hand, consider the action of matter $S_m$ as a combination of a scalar field $\phi(x)$ 
  and a nonlinear electromagnetic field minimally coupled to gravity, 
\beq                       \label{S_m}
		S_m = \int \sqrt{-g}\, d^4x \big[ 2 h(\phi) g\MN \d_\mu\phi \d_\nu\phi - 2V(\phi) - \cL(\cF) \big],
\eeq
  where the function $h(\phi)$ realizes the scalar field parametrization freedom: substituting 
  $\phi = \phi(\tilde\phi)$, we obtain another function ${\tilde h}(\tilde\phi)$ having, however, the same sign 
  as $h(\phi)$. In particular, if $h(\phi) >0$, we are dealing with a canonical scalar field with positive 
  kinetic energy, while $h(\phi) < 0$ corresponds to a phantom scalar field having negative kinetic energy. 
  A reparametrization then makes it possible to get $h(\phi) \equiv 1$ for a canonical field and 
  $h(\phi) \equiv -1$ for a phantom one. If $h(\phi)$ somewhere changes its sign, it means that the 
  $\phi$ field changes its nature from one range of its values to another, as happens, for example, in 
  wormhole models with a trapped ghost, where the scalar is canonical in a weak field region and phantom 
  in a strong field one \cite{trap10}.  

  Variation of the action \rf{S_m} with respect to the metric tensor $g^{\mu\nu}$ leads to the SET
\beq                     \label{T-fields}
               T\mN = T\mN[\phi] + T\mN[F],
\eeq
  with
\bearr                   \label{SET-phi}
			T\mN[\phi] = 2 h(\phi) \d_{\mu}\phi\d^{\nu}\phi 
				- \delta\mN \big[ h(\phi) g^{\rho\sigma}\d_\rho \phi \d_\sigma\phi -V(\phi) \big],
\yyy                    \label{SET-F}
			T\mN[F] = - 2 \mathcal{L_F} F_{\mu\sigma} F^{\nu\sigma} 
					+\frac 12 \delta\mN \mathcal{L(F)},
\ear 
  where $\mathcal{L_F} = d\cL/d\cF$. Variation of $S_m$ in $\phi$ and the \emag\ potential $A_\mu$
  gives the field equations
\bearr         \label{eq-phi}
			2 h(\phi) \nabla_{\mu}\nabla^{\mu}\phi 
					+ \frac{d h}{d\phi} \phi^{,\mu}\phi_{,\mu} + \frac{d V(\phi)}{d \phi} =0,
\yyy		\label{eq-F}
			\nabla_\mu(\mathcal{L_F}F\MN) = 0.
\ear

  The space-time symmetry encoded in \rf{ds} imposes evident requirements on the scalar and \emag\ fields as 
  possible sources of the metric. Specifically, we put $\phi = \phi(x)$ and single out the possible nonzero 
  components of $F\mn$ compatible with spherical symmetry, that is, $F_{tx}= -F_{xt}$ (a radial electric field) 
  and $F_{\theta\varphi} =-F_{\varphi\theta}$ (a radial magnetic field). Let us assume that there is only
  a magnetic field, with $F_{\theta\varphi} = -F_{\varphi\theta}= q \sin\theta$, where $q$ is a monopole 
  magnetic charge. Such a form of $F\mn$ is common for any choice of $\cL(\cF)$.
  In this case, \eq \rf{eq-F} is trivially satisfied, and the \emag\ invariant $\cF$ takes the form 
  $\mathcal{F} = 2 q^2/r^4$. As a result, the SETs \rf{SET-phi} and \rf{SET-F} read
\bearr         \label{T-phi}
		T\mN[\phi] = h(\phi) A(x) \phi'{}^2 \diag (1, -1, 1, 1) + \delta \mN V(\phi), 
\yyy           \label{T-F}		
		T\mN[F] = \frac 12 \diag\Big(\cL,\ \cL,\ \cL - \frac{4q^2}{r^4} \cL_\cF,\
				\cL - \frac{4q^2}{r^4} \cL_\cF\Big). 
\ear  

  It is clear that a scalar or \emag\ field taken separately cannot account for an arbitrary metric \rf{ds} 
  because the quantities $\rho, p_r, p_\bot$, obtained by substituting \rf{ds} to the Einstein equations, 
  are in general all different, whereas the SET $T\mN[\phi]$ has the property $T^t_t = T^\theta_\theta$, 
  and $T\mN[F]$ has the property $T^t_t = T^x_x$. 
  More than that, the same relation $T^t_t = T^\theta_\theta$ holds for more general scalar 
  fields minimally coupled to gravity, e.g., with any Lagrangian of the form $L(\phi, X)$, 
  $X:= \d_\mu \phi \d^\mu \phi$ (called generalized k-essence fields), so such fields are not more suitable 
  for our purpose than $\phi$ in \eq \rf{S_m}. However, taken together, scalar and \emag\ fields can 
  provide a source for any metric \rf{ds}.
  
  Indeed, if we know the functions $A(x)$ and $r(x)$, we also know $\rho, p_r, p_\bot$ as functions of $x$ 
  and can try to identify them with the corresponding components of the SET \rf{T-fields}. Then we immediately 
  obtain that 
\beq              \label{T02}
  		T^t_t - T^\theta_\theta = \frac{2q^2}{r^4} \cL_\cF = \rho + p_\bot, 
\eeq
  i.e., we know $\cL_\cF$ as a function of $x$, and since $\cF(x) = 2q^2/r^4(x)$ is also known, the functions 
  $\cL(x)$ and finally $\cL(\cF)$ can be calculated, at least in ranges where $r(x)$ is monotonic. Furthermore,
\beq              \label{T01}
		T^t_t - T^x_x = 2 h(\phi) A(x) \phi'{}^2 = \rho + p_r,
\eeq   
  which gives us $h(\phi)\phi'{}^2$ as a function of $x$. One can notice that if $\rho + p_r \geq 0$, in 
  accord with the Null Energy Condition (NEC), we must take $h(\phi)\geq 0$ and can put $h(\phi) \equiv 1$ 
  without loss of generality, and the scalar field is canonical. If $\rho + p_r \leq 0$, violating the NEC, 
  we can safely put $h(\phi) \equiv -1$, corresponding to a phantom scalar field. If $\rho + p_r$ changes its
  sign, then, simultaneously, the sign of $h(\phi)$ must also change, and there remains a freedom to 
  choose the scalar field parametrization. Thus we find out $\phi(x)$ and $h(x)$. 
  To calculate the only quantity still remaining unknown, the potential $V(\phi)$, we can use any of the 
  components of \eqs \rf{EE}, for example, $G^t_t = - T^t_t$. Since the scalar field equation \rf{eq-phi} 
  is a consequence of the Einstein equations, the whole set of equations is thus fulfilled. 

  This algorithm was described in a more special setting in \cite{rahul22} and applied to Simpson-Visser 
  space-times that regularize the \Scw\ and \RN\ metrics. Further on in this paper we will consider some other 
  well-known space-times containing \bhs\ or naked singularities, formulate their regularization by analogy 
  with the Simpson-Visser suggestion and try to determine their possible scalar-NED sources.  
  
  In fact, this algorithm does not depend on a particular choice of the radial coordinate, but in the examples 
  to be considered we will use the form \rf{ds} of the metric, for which the nontrivial components of 
  the Einstein equations read 
\bearr           \label{EE0}
  	 G^t_t = \frac{1}{r^2} [-1 + A (2 r r'' + r'{}^2) + A' r r'] = - T^t_t , 
\\ \lal  	        \label{EE1}
	 G^x_x = \frac{1}{r^2} [-1 +  A' r r' + A r'{}^2] = - T^x_x , 
\\ \lal           \label{EE2}
	 G^\theta_\theta = G^\varphi_\varphi - \frac{1}{2r} [2 A r'' + r A'' + 2 A' r'] = - T^\theta_\theta
\ear   
  (the prime stands for $d/dx$), and in particular, 
\beq 			\label{EE01}
		G^t_t - G^x_x =  2A(x) \frac{r''}{r} = - (T^t_t - T^x_x). 
\eeq  

\section{Regularized Fisher space-time}

  As our first example, let us consider the \ssph\ solution to the Einstein equations with a canonical massless
  scalar field $\Phi$, first obtained by I.Z. Fisher in 1948 \cite{fisher48} and a few times rediscovered later on
  (e.g., by Janis, Newman and Winicour in 1968 \cite{JNW68}, so that it is sometimes called the JNW solution). 
  This solution corresponds to the source \rf{S_m} with $h(\phi) \equiv 1$, $V \equiv 0$, without an 
  \emag\ field, and can be written as (see, e.g, \cite{BR-book})\footnote
  		{At $a=1$ this metric restores the \Scw\ solution with mass $m = k$. At $a > 1$, the same metric 
  		belongs to one of the subfamilies of the so-called anti-Fisher solution to the Einstein equations with a 
  		massless phantom scalar, \eqn{S_m} with $h(\phi) \equiv -1$ and also without an \emag\ field
  		\cite{afish57, kb73, BR-book}. The space-time \rf{fish} then has a throat at some $x > 2k$, and 
  		$r \to \infty$ as $x \to 2k$, but this subfamily does not contain \whs\ since $A(x)\to 0$ as $x\to 2k$, 
  		which is in general a singularity, except for some special cases comprising ``cold \bhs'' \cite{kb-cold}
  		with infinite horizon area and zero Hawking temperature.
  		}
\bearr             \label{fish}
		ds^2 = \Big(1- \frac{2k}{x}\Big)^a dt^2 - \Big(1- \frac{2k}{x}\Big)^{-a} dx^2 
					- x^2 \Big(1- \frac{2k}{x}\Big)^{1-a} d\Omega^2,
\nnn
	    \Phi = \pm \frac{\sqrt{1-a^2}}{2} \ln \Big(1- \frac{2k}{x}\Big), 
\ear    
  where $k > 0$ and $a \in (-1, 1)$ are integration constants, such that $m = ak$ has the meaning of the 
  \Scw\ mass, and $C = \pm k\sqrt{1-a^2}$ is a scalar charge.  
    
  In the solution \rf{fish}, $x$ ranges from $2k$ to infinity, and $x=2k$ is a naked singularity. Therefore, to 
  regularize the metric in the spirit of Simpson and Visser's proposal, one can replace the difference 
  $x - 2k = y$ with the expression $\sqrt{u^2 + b^2}$, where $u$ is a new radial coordinate, ranging in 
  $u \in \R$, and $b > 0$ is a new constant with the dimension of length. It results in the metric       
\beq              \label{fish-reg}
		ds^2 = \Big(\frac{y}{y+2k}\Big)^a dt^2 - \Big(\frac{y}{y+2k}\Big)^{-a} du^2 
					- y^{1-a} (y+2k)^{1+ a} d\Omega^2,   \qq  y = \sqrt{u^2 + b^2},
\eeq
  which is regular at all $u \in \R$ and \asflat\ at $u \to \pm\infty$. It is easy to see that \rf{fish-reg} describes
  a static traversable \wh\footnote 
  		{Another ``deformation'' of the metric \rf{fish} was recently considered in \cite{k_pal22}, in the 
  		present notations it corresponds to $x = \sqrt{u^2 + b^2}$. In such a case, to get rid of the 
  		singularity at $x = 2k$, one has to require $b > 2k$.} 
  with a \Scw\ mass equal to $ak$ at both flat asymptotics $u \to \pm \infty$ and a throat at $u=0$ with 
  the radius
\beq             \label{fish-th}
			r_{\rm th} = b^{(1-a)/2} (b+2k)^{(1+a)/2}.
\eeq  
  However, this metric is no more a solution of GR with a massless scalar field, and its possible source can 
  be found as outlined in the previous section. 
  
  Accordingly, with matter described by the action \rf{S_m}, the difference between \rf{EE0} and \rf{EE1} yields
\beq                        \label{01-fish}
			\frac{r''}{r} = -h(\phi) \phi'{}^2,\qq        r(u) = y^{(1-a)/2} (y+2k)^{(1+ a)/2},
\eeq  
  where the prime means $d/du$. Explicitly, 
\beq                        \label{r''-fish}
		\frac{r''}{r} = \frac{y^2 [b^2 - k^2 (1-a^2)] +b^2[k^2(1-a)(3+a) +(3-a) ky]}{y^4(2k+y)^2}
\eeq  
  (recall that $y = \sqrt{u^2+b^2}$). Equation \rf{01-fish} allows for finding $\phi(u)$ if $h(\phi)$ is known.
  However, in general, $h(\phi)$ should change its sign together with $r''/r$, therefore, it makes sense to use 
  the parametrization freedom of the scalar field and simply choose for it some monotonic function in the 
  range $u \in \R$, for example,
\beq                                \label{phi1}
			\phi(u) = \arctan (u/b) \ \ \Longleftrightarrow \ \ u = b \tan \phi,\qq   \phi\in(-\pi/2, \pi/2).
\eeq  
  Then we determine $h(\phi)$ from \rf{01-fish}, with the result
\beq  					\label{h1}
		   h(\phi) = - \frac{y^2 [b^2 - k^2 (1-a^2)] +b^2[k^2(1-a)(3+a) +(3-a) ky]}{b^2 (2k+y)^2},
\eeq   
  where, according to \rf{phi1}, we should substitute $y = b/\cos\phi$. At $u\to 0$ and $u\to \pm\infty$,
  $h(\phi)$ has the following limits:
\beq
			h(\phi)\Big|_{u\to 0} \to - \frac{2k^2 (1-a)+ b^2 + bk(3-a)}{(b+2k)^2}, \qq  
			h(\phi)\Big|_{u\to \pm\infty} \to \frac{k^2(1-a)-b^2}{b^2}.  
\eeq  
  As was expected, at $u=0$ (the throat), we have $h < 0$, corresponding to a phantom scalar field.
  However, at large $u$ the sign of $h$ depends on the parameters of the model (see Fig.\,1): at large $b$ we 
  have also $h< 0$, so that the scalar remains phantom in the whole range of $u$ (and $\phi$), but at small 
  $b$, such that $b^2 < k^2(1-a)$, the scalar field at infinity is canonical, so that we deal with the so-called
  trapped ghost scalar that has phantom properties only in a strong field region \cite{trap10}, see Fig.\,1.
\begin{figure} \centering   
\includegraphics[scale=0.4]{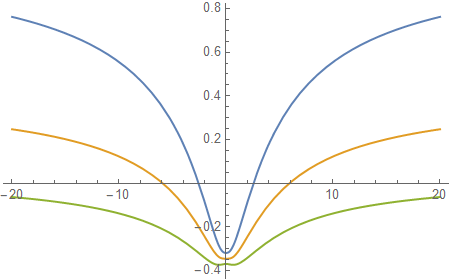}    \qq                    
\includegraphics[scale=0.3]{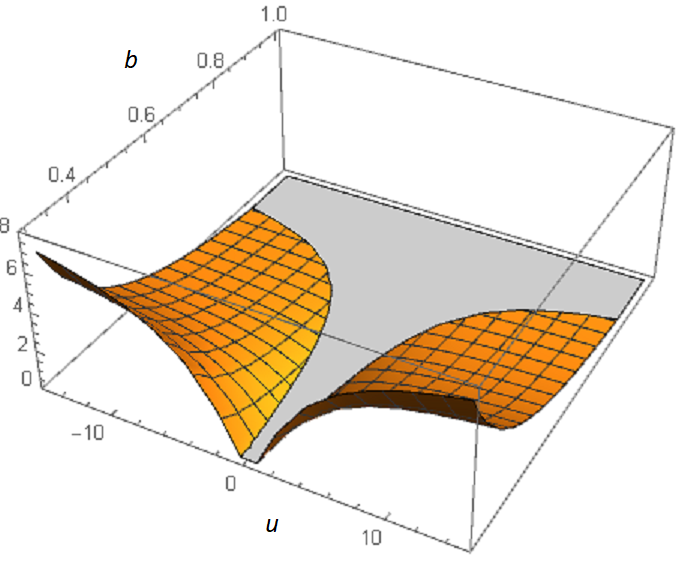}
\caption{\protect\small 
		The function $h(\phi(u))$, \eqn{h1}. Left: $k=1,\ a=0.7, \ b= 0.5,\ 0.6,\ 0.7$ (upside down). 
		Right: $k=1, \ a=0.3$, the gray ``floor'' indicates areas where $h(\phi) <0$. At other values of $a$ 
		the picture is qualitatively similar: $h < 0$ near the throat ($u=0$) and remains negative in all space
		at sufficiently large $b$ but changes its sign and thus exhibits a ``trapped ghost'' behavior at 
		smaller $b$.   
		}                      
\end{figure}                                           
  
  Our next task is to determine the suitable NED Lagrangian $\cL(\cF)$ from the difference of \eqs \rf{EE0}
  and \rf{EE2}. A calculation gives
\beq
			\cF \frac {d\cL}{d\cF} = G^\theta_\theta - G^t_t 
				= \frac {b^2 k (2a-1)}{y^{4-a}(2k+y)^{1+a}}.
\eeq  
  Since $\cF = 2q^2 /r^4$ and $r^2 = y^{1-a} (2k+y)^{1+k}$, it is easy to find $d\cF/\cF$ in terms of
  $y$ (and it is now not important that we have put $y = \sqrt{u^2 + b^2}$):
\beq
			\frac{d\cF}{\cF} = - \frac{4[y + k(1-a)]}{y (2k+y)},
\eeq   
  and therefore we can calculate $\cL$ as a function of $y$ by integrating the expression 
\beq
			\frac {d\cL}{d y} = \frac {4 b^2 k (1-2a) [y + k(1-a)]}{y^{5-a}(2k+y)^{2+a}},
\eeq  
  with the result
\bearr                     \label{L1}
       \cL(\cF) = \frac{(2 a - 1) b^2 (1 + 2 k/y)^{-a}}{(4 - a) (3 - a) (2 - a) (1 - a) a k^4 y^4 (2k+y)}
        \bigg\{4 a^5 k^5 - 9 [-1 + (1 +2 k/y)^a] y^4 (2 k + y)
\nnnv    \cm   \ \ 
        + a^4 (-28 k^5 + 6 k^4 y) + 4 a^3 k^3 (17 k^2 - 3 k y + 3 y^2)
         -  2 a^2 (34 k^5 + 3 k^4 y - 9 k^2 y^3) 
\nnn	 \inch
         +  6 a k (4 k^4 + 2 k^3 y - 2 k^2 y^2 + 3 k y^3 + 3 y^4) \bigg\},
\ear    
  where, to get an explicit function of $\cF$, one should substitute $y = y(\cF)$ as a solution to the
  transcendental equation  $y^{2-2a} (y+2k)^{2+2a} = 2q^2/\cF$.
  
  A feature of interest is that $\cL(\cF) =0$ in the case $a = 1/2$, which means that the regular metric
  \rf{fish-reg} is obtained with only a scalar source, without NED. In fact, it becomes possible because in 
  this case we have $G^t_t = G^\theta_\theta$ for the metric \rf{fish-reg}.
  
  In the general case, the asymptotic behavior of  $\cL(\cF)$ at large $|u|$ is 
\beq
		 \cL(\cF) \approx \frac{4(2a -1) b^2 k}{5 |u|^5}\ \ {\rm as} \ \ u \to \pm \infty,
\eeq 
    A more general picture is illustrated in Fig.\,2. 
\begin{figure} \centering   
\includegraphics[scale=0.3]{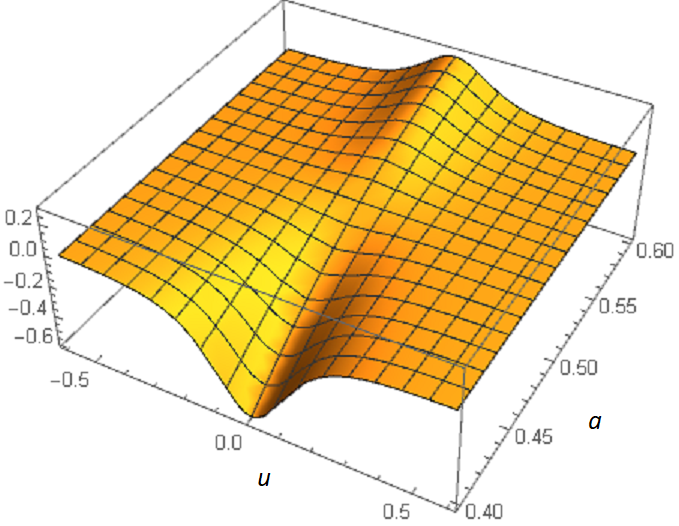}   
\caption{\protect\small 
		The function $\cL(\cF(u))$ at $k=1,\ b=0.1$. Its behavior is similar at other values of $b$: it is
		regular and smooth, it is positive at $a > 1/2$, zero at $a=1/2$ and negative at $a <  1/2$. 
		}                      
\end{figure}                                           
\begin{figure*} \centering   
\includegraphics[scale=0.5]{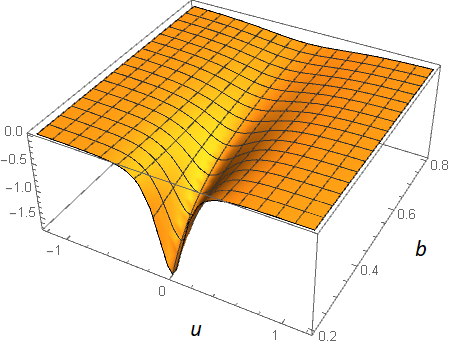}  
\caption{\protect\small 
		The potential $V(\phi(u))$ for the scalar field \rf{phi1}, \rf{h1}, with $k =1,\ a = 0.5$ 
		(exceptional case). The behavior of $V(\phi)$ turns out to be qualitatively
		the same in the exceptional and generic cases.		}                      
\end{figure*}                                           
    
  The last quantity to be determined is the potential $V(\phi)$, which can be found, for example, from 
  the radial component \rf{EE1} of the Einstein equations,
\beq                     \label{V}
			V(\phi) = -G^u_u + A(u) h(\phi) \phi'{}^2 - \Half \cL(\cF). 
\eeq  
  where we must substitute $G^u_u$ for the metric \rf{fish-reg}, $A(u)$ from the same metric, $h$, 
  $\phi$ and $\cL(\cF)$ as determined above. The calculation gives
\beq                    \label{V1}
   		    V(\phi) = - \frac{(1-a) b^2 k} {y^{4-a} (2 k+y)^{1+a}} - \Half \cL(\cF),
\eeq  
  where $\cL(\cF)$ is given by \rf{L1}.
  In the exceptional case $a = 1/2$ where the regularized metric \rf{fish-reg} is sourced by the scalar field 
  $\phi$ alone, its potential has the form
\beq
			V(\phi)\bigg|_{a=1/2} =- \frac{b^2 k} {2 y^{7/2} (2k + y)^{3/2}},
\eeq  
  where the substitution $y = b/\cos\phi$ leads to an explicit expression in terms of $\phi$. A more general 
  qualitative behavior of $V(\phi)$ is shown in Fig.\,3. At large $|u|$ we have $V(\phi)\sim 1/u^5$. 

\section{Regularized dilatonic black hole}

  Dilatonic \bhs\ are space-times obtained as special solutions to the Einstein equations with a material source 
  representing of a massless scalar field interacting with an electromagnetic field as described by the action 
\beq              \label{S-dil}
		S_{\rm dil} = \Half \int \sqrt{-g}\, d^4x \big[2 g\MN \Phi_{,\mu}\Phi_{,\nu}
						- \e^{2\lambda\Phi} F\mn F\MN \big],									 
\eeq  
  where $\lambda$ is a coupling constant. The special solution to be considered can be written with the metric 
  \rf{ds} such that \cite{dil1, dil2, dil3, dil4}
\beq                                       \label{ds-dil}
		A(x) = \Big(1 - \frac{2k}{x}\Big)\Big(1 + \frac px \Big)^{-2/(1+\lambda^2)},
		\qq
				r^2(x) = x^2 \Big(1 + \frac px \Big)^{2/(1+\lambda^2)},
\eeq
  with the scalar ($\Phi$) and electric ($\vec E$) fields given by
\beq               \label{phi-dil}
			\Phi = - \frac {\lambda}{1+\lambda^2} \ln \Big(1+ \frac px \Big),
			\qq
			2{\vec E}{}^2 = -F\mn F\MN = \frac {Q^2}{r^4(x)} \e^{-4\lambda \Phi},  
\eeq  
  where $k > 0$ and $Q$ (the electric charge) are integration constants, and
  $p = \sqrt{k^2 + Q^2(1+\lambda^2)} - k > 0$. 

  Let us focus on the case $\lambda = 1$, related to string theory \cite{dil2, dil3}. The metric takes the simple form
\beq                  \label{ds-dil1}
		ds^2 = \frac{1-2k/x}{1+p/x} dt^2 - \frac{1+p/x}{1 -2k/x} dx^2 - x (x+p) d\Omega^2.
\eeq
  This space-time has the \Scw\ mass $m = k + p/2 = Q^2/p$, a horizon at $x = 2k$, and a singularity 
  at $x =0$. The global causal structure is the same as that of the \Scw\ space-time. 
  
  As before, let us regularize this space-time by replacing $dx \mapsto du$ and $x \mapsto \sqrt{u^2 + b^2}$,
  $b>0$:
\beq                \label{ds-dil-u}
		ds^2 = \frac{1-2k/x}{1+p/x} dt^2 - \frac{1+p/x}{1 -2k/x} du^2 - x (x+p) d\Omega^2, 
		\qq    x = \sqrt{u^2 + b^2}.
\eeq
  Evidently, the range of $u$ is $u \in \R$, the metric \rf{ds-dil-u} is \asflat\ at $u\to\pm \infty$ and describes:
    
  (i) if $b < 2k$, a regular \bh\ with two horizons at $u = \pm \sqrt {4k^2 - b^2}$ and a black bounce 
  at $u=0$; 

  (ii) if $b = 2k$, a regular extremal \bh\ with a single extremal horizon (a black throat \cite{rahul22}) at $u=0$;

  (iii) if $b > 2k$, a symmetric traversable \wh\ with a throat at $u=0$; the throat radius is 
  $r_{\rm th} = \min r(u) = \sqrt{b (p +b)}$.
  
  The metric \rf{ds-dil-u} is not a solution of GR with matter specified by \rf{S-dil} but should be a solution
  corresponding to \rf{S_m}. Let us determine its particular form. As in the previous section, we can begin with 
  the scalar $\phi$, and quite similarly to \eqs \rf{01-fish}--\rf{h1}, we now have  
\bearr                        \label{01-dil}
			r(u) = \sqrt{x(x+p)},\qq  
			h(\phi) \phi'{}^2 = - \frac{r''}{r} =
					- \frac{(4 b^2 -p^2) x^2 +6 b^2 px + 3b^2 p^2}{4 x^4 (x+p)^2},			
\ear 
  (recall that $x = \sqrt{u^2+b^2}$, and the prime means $d/du$). Furthermore, using the parametrization 
  freedom of $\phi$, we put again
\bearr                                \label{phi2}
			\phi(u) = \arctan (u/b), \qq   u = b \tan \phi, \qq   
\yyy                                    \label{h2}
		 \then \   h(\phi) = - \frac{(4 b^2 -p^2) x^2 +6 b^2 px + 3b^2 p^2}{4 b^2 (x+p)^2},
\ear
  where, according to \rf{phi2}, we should substitute $x = b/\cos\phi$. As $u\to 0$ and $u\to \pm\infty$,
  $h(\phi)$ behaves as follows:
\beq
			h(\phi)\Big|_{u\to 0} \to - \frac{2b + p}{2 (b+p)}, \qq  
			h(\phi)\Big|_{u\to \infty} \to  \frac{p^2 - 4 b^2}{4b^2}.  
\eeq    
   At $u=0$ we have $h < 0$, corresponding to a phantom $\phi$ field. At large $u$, it turns out that
   the $\phi$ field is canonical ($h >0$) at $b < 2p$ and phantom at larger values of $b$. Thus at sufficiently 
   small values of the regularizing parameter $b$, we again meet a ``trapped ghost'' scalar as a source 
   of the geometry, see Fig.\,4.
\begin{figure*} \centering   
\includegraphics[scale=0.4]{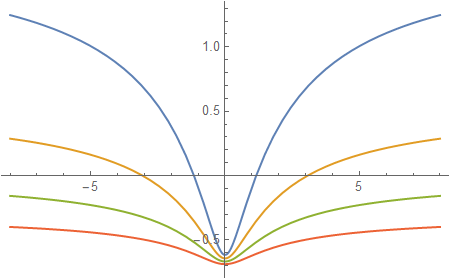}       \qq                
\includegraphics[scale=0.4]{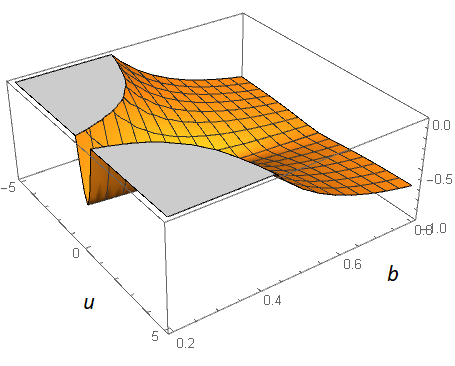}        
\caption{\protect\small 
		The function $h(\phi(u))$, \eqn{h2}, with $p=1,\ b= 0.3,\ 0.4,\ 0.5,\ 0.6$ (left, upside down),
		and for a range of $b$ (right), where gray tops show regions with $h(\phi(u)) > 0$. At 
		sufficiently small values of $b$ at given $p$, the scalar $\phi$ has a trapped-ghost nature.  
		}                      
\end{figure*}                                           

  Next, the difference of \eqs \rf{EE0} and \rf{EE2} allows for finding $\cL(\cF)$:
\bearr                   \label{L2'}
		\frac{d\cL}{dx} = \frac 1\cF \frac{d\cF}{dx} (G^\theta_\theta - G^t_t)
			= -\frac{2 (p+2 x)}{x^5 (p+x)^4} 
				\Big(b^2 \left[k (p^2+2 p x+3 x^2)+p x^2\right]+p (2 k+p)x^3 \Big),
\yyy                 \label{L2}
		\cL(\cF) = \frac 2{p^5} [b^2 (k+2 p)-2 p^2 (2 k+p)] \ln \frac {x+p}{x}
\nnn	\cm	
		- \frac{1}{6 p^5 x^4 (p+x)^3}  \Big\{ - 4 p^3 x^3 (2 k+p) (3 p+2 x)(p^2+3 p x+3 x^2)
\nnn		\cm
		+ b^2 p \big[-3 k p^6-9 k p^5 x - 3 p^4 (5 k+2 p) x^2 + (k+2 p)(3 p^3 x^3 +22 p^2 x^4 
		+30 p x^5 + 12 x^6) \big] \Big\},
\ear				
  where, to really obtain $\cL$ as a function of $\cF$, one should substitute $x$ as a solution to the 
  quartic equation $x^2 (x+p)^2 = 2q^2/\cF$. Thus the expression of $\cL(\cF)$ is quite complicated.
  Still it is clear that it is a regular function, and we illustrate its behavior in terms of $x$ for some examples 
  of the parameter values in Fig.\,5. 
\begin{figure} \centering   
\includegraphics[scale=0.4]{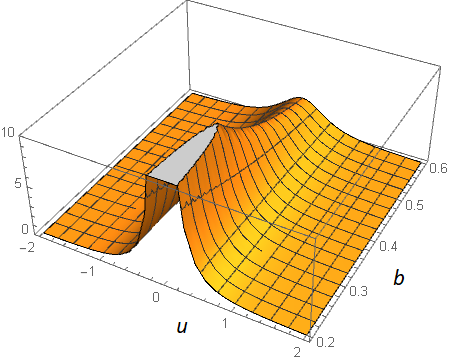}                        
\caption{\protect\small 
		The NED Lagrangian function $\cL(\cF(u))$, \eqn{L2}, at $k=p=1$.
		At other values of the parameters the behavior of  $\cL(\cF(u))$ is qualitatively the same.
		}                      
\end{figure}                                           
  Note that at large $x$ we have 
\beq                                 \label{F2}
  		\cL(\cF) \approx  \frac{p(2k+p)}{x^4} \approx  \frac{2Q^2}{r^4} = \frac {Q^2}{q^2}\cF,
\eeq
  it exhibits a Maxwell asymptotic behavior. Recall that $Q$ is the electric charge in the dilatonic \bh\
  \rf{ds-dil1}, \rf{phi-dil} with $\lambda =1$, which has nothing to do with the magnetic charge $q$ used 
  in the NED model that supports the regularized solution.
\begin{figure} \centering   
\includegraphics[scale=0.35]{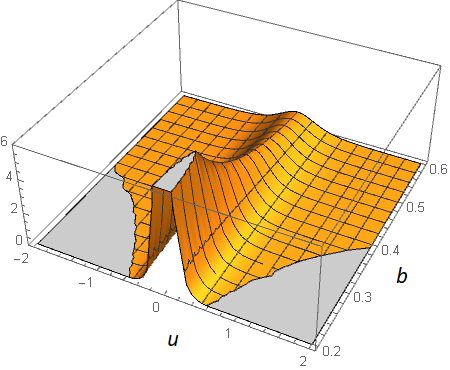}    \ \                    
\includegraphics[scale=0.35]{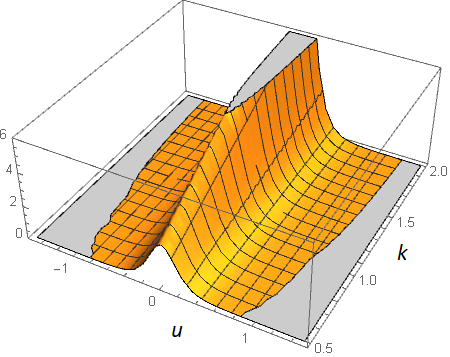}   \ \
\includegraphics[scale=0.35]{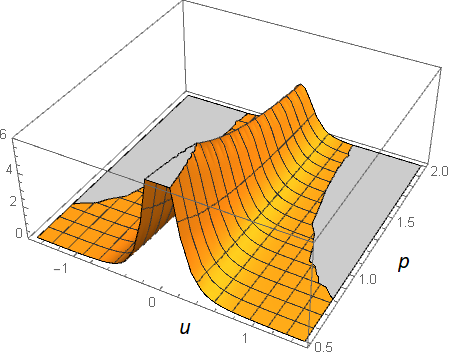}
\caption{\protect\small 
		The potential $V(\phi(u))$, \eqn{V2}, for the scalar field \rf{phi2}, \rf{h2} with 
		$k = 1,\ p=1$ and different $b$ (left),
		$p = 1,\ b =0.3$, and different $k$ (middle), and
		$k = 1,\ b =0.3$, and different $p$ (right).
		Gray bottoms show the regions where $V(\phi(u)) < 0$.
		}                      
\end{figure}                                           
  
  The scalar field potential $V(\phi)$ is again obtained according to \rf{V} and reads
\bearr                      \label{V2}
		V(\phi) = 
				\frac{p (2 k + p) x^3 + 
			 b^2 \big[p x^2 + 2 k (p^2 + 2 p x + 2 x^2)\big]}{2 x^4 (p + x)^3} - \Half \cL(\cF),
\ear
  where $\cL(\cF)$ is given by \rf{L2}, and, as before, $x = b/\cos\phi$. At large $x$ we have 
  $x \approx |u| \approx r(u)$, and $V(\phi)$ behaves as 
\beq
			V(\phi) \approx \frac{8 b^2 k + b^2 p - 2 k p^2 - p^3}{10 |u|^5}.
\eeq   
  The behavior of $V$ in the whole space is illustrated in Fig.\,6.
  
\section{Concluding remarks}
  
  In the previous studies of \sph\ black-bounce space-times, the SV regularization trick was applied to
  the spherical radius $r$ in the form $r = \sqrt{u^2+b^2}$, even though some general reasoning used 
  an arbitrary function $r(u)$ \cite{lobo20, simp21} (or $\Sigma(r)$ in their notation).  Consequently, 
  the expression $r''/r = b^2/r^4$, determining the sign of $h(\phi)$ (see \rf{T01} and \rf{EE01}), 
  is everywhere positive, making $h(\phi) < 0$ in the whole space, and a scalar field $\phi$, able to support
  the corresponding regular metric, is inevitably phantom. Unlike that, in our examples \rf{fish} and 
  \rf{ds-dil1} it is more reasonable to make the corresponding replacement not in $r$ but in a parameter 
  whose zero value leads to a singularity. As a result, $r''/r $ is, in general, not everywhere nonnegaitive, 
  and the field $\phi$ supporting the model is then necessarily of trapped-ghost nature, which is a new
  feature of this kind of models.
  
  It is not surprising that a general \ssph\ metric \rf{ds}, containing two arbitrary functions $A(x)$ and $r(x)$ 
  can be supported by a matter source with also two arbitrary functions, $V(\phi)$ and $\cL(\cF)$, but the
  regularization is here slightly complicated by the necessity of trapped-ghost fields. 
  
  As to the NED source of the same models, it cannot affect the NEC violation related to $r''/r >0$ due to the 
  equality $T^t_t = T^x_x$, see \rf{SET-F} and \rf{T-F}. If this equality holds for a full SET, 
  a regular minimum of $r(x)$ is impossible, therefore \ssph\ \whs\ with a purely NED source cannot 
  exist: there can be either magnetic black holes or solitons with a regular center or dynamic \whs\ 
  existing in a finite period of time, see, e.g., \cite{kb01-NED, kb18-NED} and references therein. 
  
  By construction, the regularized configurations considered here are $\Z_2$-symmetric with respect to the 
  minimum-$r$ sphere $u=0$.\footnote
  		{Though, some kinds of nonsymmetric regularizations were considered in \cite{lobo20}.} 
  Therefore, \bhs\ with a single horizon, like the \Scw\ one or the dilatonic one
  given by \rf{ds-dil}, turn into regular \bhs\ with two horizons (at least for small values of the regularization 
  parameter $b$); \bhs\ with two horizons like the \RN\ ones turn into those with four horizons, etc. Thus 
  the regularization substantially complicates the global causal structure of space-times, as demonstrated,
  in particular, by Carter-Penrose diagrams for three- and four-horizon \bhs\ presented in \cite{rahul22}
  and occupying the whole plane plus a countable set of overlappings.  It is also clear that thus 
  regularized \bh\ metrics cannot have any kind of scalar field as their only source since 
  it would violate the global structure theorem \cite{kb01-glob} from which it follows that an \asflat\ \ssph\
  configuration in GR with a scalar source cannot contain more than one horizon. Unlike that, regularization 
  of a metric with a naked singularity leads to a \wh\ whose source can be a scalar field alone, 
  and such an example has been really obtained here with the metric \rf{fish} in the special case $a = 1/2$. 
  A single scalar field source for an arbitrary metric \rf{ds} can be obtained (though under some 
  resrictions) if we consider scalar-tensor theories instead of GR \cite{kb23}, due to arbitrariness of the
  nonminimal coupling function $f(\phi)$ in the Lagrangian. 
  
  A problem of interest is the stability of regularized space-times, and it is important to mention that 
  the stability properties of a given geometry can be different, depending on the dynamics of the sources
  of this geometry. For example, the simplest Ellis wormhole \cite{ellis73, kb73} can be stable or unstable 
  depending on the nature of its source --- a phantom scalar, a kind of perfect fluid or a k-essence field 
  \cite{gonz08, kb-sha13, kb-fa21}. We can also recall that Fisher's solution \rf{fish} was found to be unstable
  due to its behavior near its naked singularity \cite{kb-hod}, and it can be a subject of a further study to 
  find out how this result may change if the singularity is replaced by a \wh\ throat and there is a 
  trapped-ghost scalar field as a source.
  
\subsection*{Acknowledgment} 
 
   I am grateful to Manuel E. Rodrigues for pointing out an error in the previous version of the paper. 


\end{document}